\documentclass[12pt,preprint]{aastex}

\shorttitle{Chemistry in evaporating ices}
\shortauthors{C. Cecchi\textendash Pestellini et al.}

\begin{document}

\title{Chemistry in Evaporating Ices ~\textendash~ Unexplored Territory}

\author{Cesare Cecchi\textendash Pestellini}
\affil{INAF ~\textendash~ Osservatorio Astronomico di Cagliari, Strada n.54, 
Loc. Poggio dei Pini, 09012 Capoterra (CA), Italy}
\email{ccp@ca.astro.it}

\author{Jonathan M. C. Rawlings}
\affil{University College London, Department of Physics and Astronomy, Gower
Street, London WC1E 6BT}
\email{jcr@star.ucl.ac.uk}

\author{Serena Viti}
\affil{University College London, Department of Physics and Astronomy, Gower
Street, London WC1E 6BT}
\email{sv@star.ucl.ac.uk}

\author{David A. Williams}
\affil{University College London, Department of Physics and Astronomy, Gower
Street, London WC1E 6BT}
\email{daw@star.ucl.ac.uk}

\begin{abstract}
We suggest that three\textendash body chemistry may occur in warm high density 
gas evaporating in transient co\textendash desorption events on interstellar 
ices. Using a highly idealised computational model we explore the chemical 
conversion from simple species of the ice to more complex species containing 
several heavy atoms, as a function of density and of adopted three\textendash 
body rate coefficients. We predict that there is a wide range of densities and
rate coefficients in which a significant chemical conversion may occur. We 
discuss the implications of this idea for the astrochemistry of hot cores.
\end{abstract}


\section{Introduction}
Hot cores are well\textendash known as sites of relatively complex chemistry in
the interstellar medium (see, e.g., \citealt{S06} for a discussion of Sgr B2
N\textendash LMH). None of these species is readily formed in conventional 
gas\textendash phase chemistry under typical conditions of interstellar clouds
nor of those pertaining in hot cores (number densities $n_{\rm H_2} 
\sim 10^7$~cm$^{-3}$, and temperatures $T \sim 200 - 300$~K). Attention has 
therefore focused on interstellar ices as a potential source of these 
relatively complex species. Laboratory experiments have shown that irradiation
of ices of similar composition by fast particles or short wavelength 
electromagnetic radiation can induce greater chemical complexity to arise in 
the ices (e.g., \citealt{O08}). Concurrently, extremely detailed computational
models by Herbst, Garrod, and their collaborators (e.g., \citealt{G08}) 
predict that a rich and relatively complex chemistry can slowly arise in the 
low temperature ices. Then, when the ices are evaporated in the very dense and
warm gas in the vicinity of a newly\textendash forming star, these complex 
molecules are released to the gas phase and can be detected as hot core 
molecules.

In this work, we suggest an alternative approach that may work in parallel 
with these chemistries to form the large organic molecules detected in hot 
cores. We ask: can three\textendash body gas\textendash phase chemistry in 
high density transient events on evaporating ices create complex molecules? In
the process of becoming a hot core, material warms up over a finite period 
of time from a low temperature (around 10 K) to about 200 K; laboratory 
experiments \citep{C04} show that desorption occurs in several distinct and 
narrow temperature bands, of which the most important for our present purposes
is the so\textendash called co\textendash desorption band. This band is when 
the major component of the ice, H$_2$O, desorbs and carries with it all other 
species. This understanding of desorption has been shown to be consistent with
current observations of hot cores \citep{V04}.
 
In the picture we present here, we assume that during the warm\textendash up of
a pre\textendash protostellar core a major part of the ices is abruptly 
converted during the co\textendash desorption event from solid phase to gas 
phase. A truly instantaneous conversion from solid to gas would create a gas 
with a number density similar to that of the solid, i.e. about 
$10^{23}$~cm$^{-3}$. However, it is unlikely that the gas attains such high 
densities, but it is nevertheless possible that the density is initially at a 
very high level indeed, although for a very short period of time. It is our 
purpose here to explore whether three\textendash body reactions in extremely 
dense and fairly warm gas can create molecules with complexity similar to 
those observed to exist in hot cores. The source species available for these 
chemical syntheses are assumed to be those molecules available in unprocessed 
interstellar ices. Can we use H$_2$O, CO, CH$_4$, etc., to create molecules of 
greater complexity in a highly transient and extremely dense phase?

Ideas of a similar kind were first explored by \citet{D00} who suggested that 
amino acids, peptides, and a variety of organometallic compounds could be 
created from evaporating ices confined within cavities inside aggregate grains.
Later \citet{CCP04} computed radiation field intensities within such cavities, 
and suggested that prebiotic molecules might arise in the chemistry promoted 
in those cavities. The processes described in these papers would take place 
over long intervals of time during the lifetime of the core gas. In this 
paper, by contrast, we consider whether the conversion from chemical 
simplicity to chemical complexity can occur very rapidly within the 
transiently very high density gas arising during the evaporation of ices in 
the co\textendash desorption phase. Of course, the processes discussed by 
\citet{D00}, by \citet{CCP04}, and by \citet{G08}, and the ideas presented 
here, are not mutually exclusive.

In Section 2 we describe the physical and chemical models that we have adopted.
The evaporating gas is assumed to be initially at a very high density (treated 
as a free parameter) and to expand either freely into a vacuum, or more slowly 
if there is some partial confinement of the gas by the dust grain geometry. The
chemistry that we adopt is hypothetical, since there is little information
available about three\textendash body gas phase chemistry at exceptionally 
high number densities. Our work, therefore, is intended to be a feasibility 
study exploring a hitherto highly transient region of parameter space. We 
present in Section 3 the results of our computational modelling. Our 
conclusions are given in Section 4.

\section{The model}
\subsection{The initial chemical conditions}
The composition of interstellar ices, measured along different 
sight\textendash lines, varies significantly (e.g., \citealt{W2nd,G04,B04}). 
We adopt a composition within the observed range. The precise composition is 
unimportant for our present purposes, where order\textendash of\textendash 
magnitude abundances of product molecules are adequate. We assume that the 
instantaneously sublimated ice mantle has a typical composition as given in 
Table \ref{tab:initial}. For our feasibility study we currently ignore other 
species, whose abundances are less than 1\% of the mantle composition. We also
do not include CO$_2$, which has a typical relative abundance of 20\%, on the 
assumption that it is tightly bound and unreactive in the assumed conditions.

\subsection{The chemical network}
With this limited species set, the likely reaction channels are also somewhat 
restricted. As the composition of the evaporating gas is dominated by H$_2$O 
and, to a lesser extent, by CO we assume that the third body in the 
three\textendash body reactions is H$_2$O.

Three\textendash body reactions are believed to have important roles in 
astronomy. They determine the chemistry in planetary atmospheres; for example,
the reaction between molecules
\[
\rm
CH_4 + H_2O + M \to CO + 3H_2 + M
\]
(where M is a third body) may affect the deep water abundance on Jupiter 
\citep{V10}. The same authors also suggest that the reaction
\[
\rm
H_2 +  H_2CO + M \to CH_3 + OH + M
\]
may open the way to radical formation (in the present work, we have 
excluded radical formation and reaction). Similarly, three\textendash body 
reactions play a role in cool stellar atmospheres \citep{T73} in the formation
of simple species such as H$_2$, CO, CO$_2$, CH$_4$, NH$_3$, C$_2$H$_2$, etc. 
Three\textendash body reactions also play a role in surface chemistry such as 
that described by \citet{G08} (see also \citealt{H92,TH82,AR77}). These surface
chemistries invoke radical/molecule reactions in which the surface is the 
third body. It should be noted that all the above chemistries rely almost 
entirely on theoretical estimates rather than laboratory determinations.

Our proposal is that three\textendash body reactions between molecules (rather
than radical/molecule reactions) may generate a product channel of 
astronomical interest. Such a channel would depend on bond\textendash breaking
and atomic rearrangement, and could be driven by the high energy tail of the 
Boltzmann distribution in the evaporating gas. Examples of this kind of 
reaction that might be possible are those quoted above for Jupiter, or 
an alternative channel
\[
\rm
CH_4 + H_2O + M \to CH_3OH + 2H + M
\]
possibly followed by
\[
\rm
CH_3OH + CO + M \to CH_3COOH + M
\]
An interesting pair of similar reactions is quoted in the NIST 
database\footnote{http://webbook.nist.gov/}
\[
\rm
H_2O + CO + M \to HCOOH + M
\]
followed by
\[
\rm
HCOOH + H_2O + M \to HOCH_2OOH + M
\]
Thus, it appears that reactions of the type we consider here are under 
consideration in various applications. We therefore assume that 
three\textendash body reactions may occur between molecular species such as 
those of the initial composition.

We do not claim to know what the products of the various reactions are; we 
only hypothesise that non\textendash defined complex organic molecules (which 
we label P$_1$ to P$_{26}$) may be formed. Obviously, this is highly 
simplistic and does not include the possibility of complex branching ratios, 
or the other (small species) products of the reactions. It therefore follows 
that we cannot track the abundances of the parent species with any degree of 
accuracy and, once the model indicates that their abundances are reduced 
through conversion into more complex molecules, then the model breaks down.

We have considered various types of possible three\textendash body reactions. 
Normally, one thinks of the process as one that involves a chemical reaction 
between two species. The third body, which is chemically inert, then 
collisionally stabilises the excited product of this reaction. However, we can 
also envisage reactions in which all three species are chemically active. The 
reactions in this category are listed in Table \ref{tab:network3}. As stated 
above, we assume that H$_2$O is always a partner in these reactions. 

The more usual formation channels would involve one of the three reactants as a
chemically passive partner. Thus to obtain the products P$_1$ to P$_{15}$ above
would require two stages of reaction, involving intermediate species (P$_{16}$ 
to P$_{26}$) after the first stage. These, first stage, reactions are listed in
Table \ref{tab:network2i}. Again, the third (now passive) partner in these 
reactions is taken to be H$_2$O.

The second stage reactions, resulting in the formation of P$_1$ to P$_{15}$, 
are given in Table~\ref{tab:network2ii}. Note that we further assume that the 
sequence of reactions to form a product is not important; thus reaction 16 
followed by reaction 27 is completely equivalent to reaction 22 followed by 
reaction 47. 

This could be extended to include reactions between less abundant species 
(e.g., CH$_4$ and OCN$^-$ etc.) but the products of these reactions would 
obviously be less abundant than the products considered here. We shall ignore 
these minor routes. 

The rate coefficients for these various reactions are entirely hypothetical 
as no detailed information is available for any of the reactions in our list.
We assume that the temperature dependences of the rate coefficients can be
described by the usual Aarhenius formalism
\begin{equation} \label{Aar}
k = k_0 \left( T/ 300~{\rm K}\right)^\alpha \exp 
 \left( -\beta/ T \right)
\end{equation}
For the purposes of this exploratory calculation we assume that all reactions
have the same basic rate coefficients $k_0$. By referring to existing 
databases of three\textendash body reactions (e.g. RATE06, \citealt{W07}) it 
is evident that most of the values of $k_0$ for the reactions lie in the 
range $10^{-27}-10^{-32}$~cm$^6$~s$^{-1}$. The values of $\alpha$ typically 
lie between 0 and $-3$, whilst the $\beta$ values are very variable, and are 
quite often negative for the temperature range for which a fit is given. We 
consider two possible values for $\alpha$: 0 and $-2$. So as to avoid 
excessive degeneracy in the parameters we set $\beta$ to zero and vary $k_0$ 
to incorporate the effects of a possible barrier. Any particular value of the 
rate coefficient can be re\textendash interpreted in terms of a canonical 
value of $k_0$ together with a particular value of $\beta$ (see Section 4).

These are the only reactions in our chemical network; at the densities that we
are considering, photoreactions and cosmic\textendash ray ionization reactions 
are entirely negligible and so have been omitted.

\subsection{The Physical Model}
We can estimate very roughly the expected timescale for significant conversion
of reactant species into complex organics on the assumptions that $(i)$ there 
are no significant activation energy barriers to the reactions in question, 
and $(ii)$ the reactions are generally constructive ~\textendash~ that is to 
say at least one the products is more complex than the reactant species. If 
the density is taken to be $n \sim 10^{23}$~cm$^{-3}$, the rate coefficients 
are all $10^{-30}$~cm$^6$~s$^{-1}$, and the fractional abundance of the 
reactants (e.g., CH$_4$, OCN$^-$ etc.) is of the order of $10^{-2}$, then the 
timescale for conversion is 
\begin{equation}
\tau \sim \frac{1}{ k \times (10^{-2} n)^2} \sim 10^{-12} \, {\rm s.}
\end{equation}
Therefore, the chemistry is very rapid indeed. So long as this timescale is 
very much less than the dynamical timescale, then the details of the expansion 
and cooling of the gas are of minimal relevance; essentially the chemistry 
takes place very shortly after the sublimation of the ice mantles when the gas 
density is highest; the relative abundances of the product molecules are then
``frozen in'' to the flow. These qualitative conclusions are confirmed by the 
detailed results presented in Section 3.

We consider a sphere of ice, instantaneously sublimated into the gas\textendash
phase, of initial radius $r_0$ and density $n_0=10^{23}$ molecules cm$^{-3}$ 
which expands into a vacuum at some fraction, $\epsilon$, of the sound speed 
$v_s$. The parameter $\epsilon$ allows for deviations from completely free, 
spherically symmetric, expansion ($\epsilon=1$) so that a value of $\epsilon=0$
would correspond to the situation of perfect trapping of the gas in a cavity.
Thus, by mass conservation
\begin{equation}
n_0 r_0^3 = nr^3 = n(r_0+ \epsilon v_s t)^3
\end{equation}
so
\begin{equation}
\frac{n}{n_0} = \left( \frac{r_0}{r_0+ \epsilon v_s t} \right)^3. 
\end{equation}
If $r_0$ is assumed to be comparable to the typical thickness of an ice mantle 
on a dust grain, then $r_0 = 10^{-5}$~cm, $v=10^4$~cm\,s$^{-1}$ and so the 
evolution of number density would in this case be given by
\begin{equation}
n/n_0 = 1 / \left( 1+10^{9}\epsilon t \right)^3 
\end{equation}
The initial temperature of the gas, $T_0$, is unknown as is the way that it 
varies with time. We are guided here by the observational results that indicate
hot core temperatures of a few hundred Kelvin. The intial temperature may be 
higher, so we examine the temperature sensitivity up to 500~K.

For an ideal gas expanding into a vacuum (Joule expansion) the temperature 
would be independent of time. For adiabatic expansion, 
$TV^{\gamma -1}=$~constant. Thus, for spherical expansion of a gas with 
$\gamma=4/3$, $T\propto r^{-1}$. In reality, the gas will cool as a result of 
work against intermolecular forces. In our model we consider limiting cases of 
\begin{equation}
T = T_0 \left( \frac{r_0}{r}\right)^q 
\end{equation}
where $q=0$ or 1. The various free parameters in our model are listed in Table 
\ref{tab:param} together with the range of values that we have investigated. 
For most of our model calculations we use the following values: $T_0=100$, 
$q=0$, $\alpha=0$, $\epsilon=1$, with $k_0$ and $n_0$ variable. We also
generally assume that the third reactant is passive ($A_3 =$~off, see
Table \ref{tab:param})

We calculate the time\textendash dependences of the chemical models using a 
simple model, based around the GEAR integration package. Although the 
three\textendash body reaction network and the extremely high densities are 
most unlike those applicable to models of molecular clouds, the principles are 
the same. As there are no variations in the chemical rate coefficients, the 
resultant sets of differential equations are not numerically stiff.

\section{Results}
The general characteristics of the results are illustrated in Tables 
\ref{tab:resultsa} and \ref{tab:resultsb}
for the case of $n_0 = 10^{21}$~cm$^{-3}$, $T_0 = 500$~K, $k_0 = 
10^{-33}$~cm$^6$~s$^{-1}$, $T \propto 1/r$ ($q=1$), $\epsilon=1$, $\alpha= -2$ 
and $A_3 =$~off. 

As can be seen from Table~6, in this model the temperature falls from its 
initial value to 200~K by about $10^{-9}$~s. The parent molecules show a slight
loss in relative abundance over this period and then remain constant. 
Conversely, we see in Table~7 that product molecules grow rapidly in relative 
abundance during a period of less than $10^{-10}$~s and achieve 
steady\textendash state within about $10^{-9}$~s. The steady\textendash
state abundances depend on the two\textendash stage chemistry and on the 
relative abundances of the parent species. We emphasise that it is assumed 
that all rate coefficients are the same and have the same temperature 
dependences, so there is no selectivity in the chemistry other than that 
introduced by the network and the initial parent abundances. In this 
particular example, products P$_1$ ~\textendash~ P$_5$ attain abundances 
relative to H$_2$O of about $10^{-4}$, products P$_6$ ~\textendash~ P$_{15}$ 
reach about $10^{-5}$, while products P$_{16}$ ~\textendash~ P$_{26}$ are in 
the range $10^{-2}-10^{-3}$. We note at this point that an abundance of 
$10^{-5}$ relative to H$_2$O may correspond to a column density in a typical 
hot core of about $10^{15}$~cm$^{-2}$. We return to this point in Section~4.

An important feature of the model whose results are given in Tables 
\ref{tab:resultsa} and \ref{tab:resultsb} is that the chemistry is dominated by
reactions occurring at the very highest densities and earliest times. 
Therefore, the relative chemistry arising from these reactions is essentially 
unchanged during the later expansion and may be said to be ``frozen\textendash
in'' to the expanding gas. The dominance of the early time in the chemistry 
(at least, for this model) also means that the dependence of the gas 
temperature on the expansion time (or radius) is fairly unimportant; this is 
also the case for the dependence of the rate coefficient on temperature. For 
example, for the case reported in Tables \ref{tab:resultsa} and 
\ref{tab:resultsb}, if all the temperature dependences are removed then the 
relative abundances are not significantly changed. In what follows, we shall 
ignore all temperature dependences.

Clearly, the assumed initial temperature, $T_0$, may influence the model 
results. In fact, a change in $T_0$ from 500~K (as in Table~2) to 100~K makes 
only rather slight changes in the relative abundances. In the model whose 
results are shown in Tables \ref{tab:resultsa} and \ref{tab:resultsb}, $k 
\propto T^{-2}$. The most significant of the remaining parameters are the 
initial density and the rate coefficients. We have therefore explored the 
density/rate 
coefficient space, over a wide range. We plot in Figure~1 a summary of these 
results for the case where at least some of the product molecules have 
relative abundances of $10^{-5}$ or larger. There is a wide range of $(n,k)$ 
space where chemistry meets this requirement for a trapping parameter of 
either 0.1 and 1.0.

For the highest density case, some chemistry occurs on timescales of 
about $10^{-10}$~s even when the rate coefficient is as low as 
$10^{-41}$~cm$^6$~s$^{-1}$; we find that products P$_{16}$ ~\textendash~ 
P$_{26}$ have relative abundances in this case of $10^{-6} - 10^{-7}$. 
We interpret the rate coefficient for these small values by assuming that 
it has a form proportional to $\exp(-\beta/T)$ (see equation \ref{Aar}), 
where $\beta$ is a barrier height. Then 
\begin{equation}
\beta = T \ln \left( k_0/k \right).
\end{equation} 
If $k_0 = 10^{-31}$~cm$^6$~s$^{-1}$ (a plausible value), then the barrier 
height for $T = 100$~K implied by a rate coefficient of 
$10^{-41}$~cm$^6$~s$^{-1}$ is 2300~K, a typical barrier height for 
bond\textendash breaking of an H\textendash X bond (see Section 4)..
For the lower densities, then the value of $k$ required to produce a 
significant chemistry is of course much larger. For example, at an initial 
density of $10^{20}$~cm$^{-3}$ with zero temperature dependence, a value 
of $k = 10^{-31}$~cm$^6$~s$^{-1}$ produces relative abundances of products 
P$_1$ ~\textendash~ P$_{15}$ of $10^{-5}-10^{-6}$. A value of 
$10^{-31}$~cm$^6$~s$^{-1}$ implies a barrier height of less than 500~K, which 
may be implausible. At an even lower initial density of $10^{17}$~cm$^{-3}$, 
then even the maximum plausible value of $k$, $10^{-29}$~cm$^6$~s$^{-1}$, 
fails to produce product molecule relative abundances larger than $10^{-6}$. 

The effect of changing the gas expansion parameter, $\epsilon$, from its 
free expansion value of 1 to a value of 0.1 that may represent a hindered 
expansion from a partially enclosed cavity, may also be considered. This 
reduction in $\epsilon$ has the effect of maintaining the gas at the highest 
density for a longer time, so that the effect is allow the chemistry to 
proceed more quickly. For example, we compare the case of gas at a 
density of $10^{22}$~cm$^{-3}$ where all rate coefficients are 
$10^{-37}$~cm$^6$~s$^{-1}$. For this case, the product molecules range over 
four orders of magnitude in relative abundances. In all cases, the relative 
abundance of a particular product molecule is larger by one order of magnitude 
when the evaporation is restricted. This behaviour is also recovered for 
a case with lower density, $10^{21}$~cm$^{-3}$ in which some products are 
present with near\textendash zero abundances. 

\section{Discussion and Conclusions}
There are a number of general conclusions that may be drawn from this 
feasibility study:

\begin{enumerate}

\item 
since the density falls rapidly in the evaporating gas, the chemistry is 
dominated by reactions occurring at the very earliest phases, typically within
about 10$^{-10}$~s, and remains ``frozen\textendash in'' at later times; if 
reactions work ~\textendash~ i.e. if activation barriers can be overcome 
~\textendash~ then the chemistry will be fast and efficient;

\item 
as a consequence, any assumed temperature evolution in the gas is unimportant, 
and any assumed temperature\textendash dependence in the adopted rate 
coefficients is also of little consequence;

\item 
therefore, the dominating parameters are the initial gas number density, $n_0$,
and the adopted three\textendash body rate coefficients, $k$;

\item 
the $(n,k)$ parameter space explored includes a large region in which the 
complex chemistry is rich enough to be observationally significant; i.e. where 
complex products have an abundance relative to water of about 10$^{-5}$, which
could be relevant for hot core chemistry. The region of $(n,k)$ parameter 
space in which this occurs is indicated schematically in Fig.1;

\item 
assuming that the form of the rate coefficients is 
\begin{equation}
k = k_0 \left( T/ 300~{\rm K} \right)^\alpha \exp \left( -\beta/ T 
\right)
\end{equation}
where $\beta$ is the barrier height in Kelvins for the assumed reactions, 
then, if $k_0$ is assumed to have a canonical value of 
10$^{-30}$~cm$^6$~s$^{-1}$, the range in $k$ of 
10$^{-30}-10^{-41}$~cm$^6$~s$^{-1}$ corresponds to a range in $\beta$ of 0 to 
2530~K if the gas temperature is about 100~K; this range in $\beta$ is 
plausible for simple bond\textendash breaking reactions, if the barrier height
is roughly about 5\% of the bond energy \citep{G41};

\item 
an important conclusion is that there is a limit to chemical complexity imposed
by the very short timescale available; for the range of parameters investigated
here, this complexity corresponds to molecules containing about three or four
heavy atoms (C, N, O); larger species are not expected from the mechanism
described here;

\item
if the ionization level is even slightly enhanced from zero (by some suitable
process, not considered here) then the rates an conversion efficiencies may be
significantly higher; similarly, if radicals were to be introduced into the 
expanding gas (a process not considered here), then a chemistry of the kind 
originally envisaged by \citet{AR77} may occur.

\end{enumerate}

The overall conclusion is this: we have demonstrated a proof of concept 
that a rich chemistry may occur in gas evaporating from a chemically\textendash
mixed ice in the co\textendash desorption event, assuming that the event is 
sufficiently narrow in temperature for very high gas densities to be achieved 
in the evaporate. A similar conclusion may be made for other desorption 
events, such as the so\textendash called 'volcano' event, but the composition 
of the evaporating gas and the products would be different.

The results from this very preliminary work are of course higly tentative, but
do seem to indicate, firstly, that there could be astronomical consequences 
from this idea, and, secondly, that further study of the idea (and of competing 
ideas involving three\textendash body chemistry) probably cannot 
be made theoretically and would require laboratory investigation.

We may crudely regard any chemistry that generates product molecules at about 
$10^{-5}$ relative to H$_2$O as potentially important from the astronomical 
point of view. Hot cores are usually assumed to have about 1000 visual 
magnitudes of extinction, corresponding approximately to a column density of 
hydrogen of about $10^{24}$~cm$^{-2}$ (e.g., \citealt{MH98}). The available 
oxygen is mainly divided between CO (formed in the gas phase in the early stage
of the gravitational collapse of the star\textendash forming core) and H$_2$O 
(formed by hydrogenation of O\textendash atoms at the surface of dust and 
retained there as the ice mantle). The column density of each species (i.e. CO
and H$_2$O) for hot cores in the Milky Way will be on the order of 
$10^{20}$~cm$^{-2}$. Therefore, molecules with an abundance relative to water 
of $10^{-5}$ may be expected to have column densities around 
$10^{15}$~cm$^{-2}$. The typical column density range of organic molecules in 
LMH is $10^{15}-10^{16}$~cm$^{-2}$ \citep{S06}; for example amino acetonitrile
(NH$_2$CH$_2$CN) with a column density of $2.8 \times 10^{16}$~cm$^{-2}$ 
\citep{B08}, and ethyl formate (C$_2$H$_5$OCHO) with a column density of $5.4 
\times 10^{16}$~cm$^{-2}$ \citep{B09}. For an extended review of complex 
organics in interstellar clouds see \citet{HvD09}. 

\citet{S06} also emphasises that those molecules are hydrogen\textendash poor. 
That seems likely to be the case for molecules arising in the processes 
considered here, as H\textendash atoms will be ejected in the bond\textendash 
breaking and bond\textendash making processes of three\textendash body 
chemistry. Indeed, the values of the rate coefficient that seem appropriate in 
this model do seem to imply bond\textendash breaking of bonds of energies a few
eV, i.e., corresponding to bonds involving H\textendash atoms. We suggest that 
an experimental investigation be made to test the validity of the ideas 
expressed here.

\acknowledgments
We  would like to thank the Royal Society for funding an exchange 
programme between UCL and Cagliari Observatory. We thank Professor S. Price and
Dr W. Brown for a very helpful discussion of the ideas in this paper.
We thank the referee for helpful comments that improved an earlier version of 
this paper.

\begin{table}
\caption{Initial gas\textendash phase composition$^{(a)}$, 
relative to H$_2$O (100) \citep{W2nd}}
\label{tab:initial}
\begin{tabular}{cc}
\\
Species & Abundance \\
\hline \hline
H$_2$O   & 100 \\
CO       & 15  \\
CH$_4$   & 4   \\
H$_2$CO  & 3   \\
CH$_3$OH & 3   \\
NH$_3$   & 1   \\
OCN$^-$  & $\lesssim 1$ \\
\hline \hline
\end{tabular}
\tablenotetext{(a)}{We do not include CO$_2$ in the chemistry, as we regard 
that species as chemically inert in this context.}
\end{table}

\begin{table}
\caption{Postulated reactions between three active partners}
\label{tab:network3}
    \begin{tabular}{cccccccc}
\\
Number & Reaction & & & & & & \\
  \hline   \hline
   1 & CO     & + & CH$_4$  & + & H$_2$O & $\to$ & P$_1$ \\ 
   2 & CO     & + & OCN$^-$     & + & H$_2$O  & $\to$ & P$_2$ \\
   3 & CO     & + & H$_2$CO & + & H$_2$O & $\to$ & P$_3$ \\
   4 & CO     & + & CH$_3$OH & + & H$_2$O & $\to$ & P$_4$ \\
   5 & CO     & + & NH$_3$  & + & H$_2$O & $\to$ & P$_5$ \\
   6 & CH$_4$ & + & OCN$^-$     & + & H$_2$O & $\to$ & P$_6$ \\
   7 & CH$_4$ & + & H$_2$CO & + & H$_2$O & $\to$ & P$_7$ \\
   8 & CH$_4$ & + & CH$_3$OH & + & H$_2$O & $\to$ & P$_8$ \\
   9 & CH$_4$ & + & NH$_3$  & + & H$_2$O & $\to$ & P$_9$ \\ 
  10 & OCN$^-$    & + & H$_2$CO & + & H$_2$O & $\to$ & P$_{10}$ \\ 
  11 & OCN$^-$    & + & CH$_3$OH & + & H$_2$O & $\to$ & P$_{11}$ \\ 
  12 & OCN$^-$    & + & NH$_3$  & + & H$_2$O & $\to$ & P$_{12}$ \\
  13 & H$_2$CO & + & CH$_3$OH & + & H$_2$O & $\to$ & P$_{13}$ \\
  14 & H$_2$CO & + & NH$_3$ & + & H$_2$O & $\to$ & P$_{14}$ \\
  15 & CH$_3$OH & + & NH$_3$ & + & H$_2$O & $\to$ & P$_{15}$ \\ 
    \hline  \hline
    \end{tabular}
\end{table}

\begin{table}
\caption{Postulated reactions between two active and one passive partner: Stage
I}
\label{tab:network2i}
    \begin{tabular}{cccccccc}
\\
Number & Reaction & & & & & & \\ 
  \hline   \hline
  16 & CO      & + &    H$_2$O & + &  H$_2$O & $\to$ & P$_{16}$ \\ 
  17 & CH$_4$  & + &    H$_2$O & + &  H$_2$O & $\to$ & P$_{17}$ \\
  18 & OCN$^-$     & + &    H$_2$O & + &  H$_2$O & $\to$ & P$_{18}$ \\ 
  19 & H$_2$CO & + &    H$_2$O & + &  H$_2$O & $\to$ & P$_{19}$ \\ 
  20 & CH$_3$OH & + &   H$_2$O & + &  H$_2$O & $\to$ & P$_{20}$ \\
  21 & NH$_3$  & + &    H$_2$O & + &  H$_2$O & $\to$ & P$_{21}$ \\
  22 & CO      & + &    CH$_4$    & + &  H$_2$O & $\to$ & P$_{22}$ \\ 
  23 & CO      & + &    OCN$^-$    & + &  H$_2$O & $\to$ & P$_{23}$ \\ 
  24 & CO      & + &    H$_2$CO   & + &  H$_2$O & $\to$ & P$_{24}$ \\ 
  25 & CO   & + &    CH$_3$OH  & + &  H$_2$O & $\to$ & P$_{25}$ \\  
  26 & CO   & + &    NH$_3$    & + &  H$_2$O & $\to$ & P$_{26}$ \\ 
  \hline    \hline
    \end{tabular}
\end{table}

\begin{table}
\caption{Postulated reactions between two active and one passive partner: Stage
II}
\label{tab:network2ii}
    \begin{tabular}{cccccccc}
\\
Number & Reaction & & & & & & \\
  \hline   \hline
  27 & P$_{16}$  & + &    CH$_4$    & + &  H$_2$O & $\to$ & P$_1$ \\ 
  28 & P$_{16}$  & + &    OCN$^-$    & + &  H$_2$O & $\to$ & P$_2$ \\ 
  29 & P$_{16}$  & + &    H$_2$CO   & + &  H$_2$O & $\to$ & P$_3$ \\ 
  30 & P$_{16}$  & + &    CH$_3$OH  & + &  H$_2$O & $\to$ & P$_4$ \\ 
  31 & P$_{16}$  & + &    NH$_3$    & + &  H$_2$O & $\to$ & P$_5$ \\ 
  32 & P$_{17}$  & + &    OCN$^-$    & + &  H$_2$O & $\to$ & P$_6$ \\ 
  33 & P$_{17}$  & + &    H$_2$CO   & + &  H$_2$O & $\to$ & P$_7$ \\ 
  34 & P$_{17}$  & + &    CH$_3$OH  & + &  H$_2$O & $\to$ & P$_8$ \\ 
  35 & P$_{17}$  & + &    NH$_3$    & + &  H$_2$O & $\to$ & P$_9$ \\
  36 & P$_{18}$  & + &    H$_2$CO   & + &  H$_2$O & $\to$ & P$_{10}$ \\
  37 & P$_{18}$  & + &    CH$_3$OH  & + &  H$_2$O & $\to$ & P$_{11}$ \\
  38 & P$_{18}$  & + &    NH$_3$    & + &  H$_2$O & $\to$ & P$_{12}$ \\
  39 & P$_{19}$  & + &    CH$_3$OH  & + &  H$_2$O & $\to$ & P$_{13}$ \\ 
  40 & P$_{19}$  & + &    NH$_3$    & + &  H$_2$O & $\to$ & P$_{14}$ \\
  41 & P$_{20}$  & + &    NH$_3$    & + &  H$_2$O & $\to$ & P$_{15}$ \\
  42 & P$_{21}$  & + &    CO     & + &  H$_2$O & $\to$ & P$_5$ \\
  43 & P$_{21}$  & + &    CH$_4$    & + &  H$_2$O & $\to$ & P$_9$ \\
  44 & P$_{21}$  & + &    OCN$^-$    & + &  H$_2$O & $\to$ & P$_{12}$ \\
  45 & P$_{21}$  & + &    H$_2$CO   & + &  H$_2$O & $\to$ & P$_{14}$ \\
  46 & P$_{21}$  & + &    CH$_3$OH  & + &  H$_2$O & $\to$ & P$_{15}$ \\
  47 & P$_{22}$  & + &    H$_2$O & + &  H$_2$O & $\to$ & P$_1$ \\
  48 & P$_{23}$  & + &    H$_2$O & + &  H$_2$O & $\to$ & P$_2$ \\
  49 & P$_{24}$  & + &    H$_2$O & + &  H$_2$O & $\to$ & P$_3$ \\
  50 & P$_{25}$  & + &    H$_2$O & + &  H$_2$O & $\to$ & P$_4$ \\
  51 & P$_{26}$  & + &    H$_2$O & + &  H$_2$O & $\to$ & P$_5$ \\
  \hline    \hline
    \end{tabular}
\end{table}

\begin{table*}
\caption{Free parameters in the model}
\label{tab:param}
    \begin{tabular}{lll}
\\
Parameter & Description & Value \\
  \hline   \hline
$k_0$ & Rate coefficient & $10^{-29}-10^{-41}$ cm$^6$~s$^{-1}$\\
$\alpha$ & Temperature dependence of rates ($k \propto T^{\alpha}$) & 
0 or $-2$ \\
$n_0$ & Initial number density & $10^{17}-10^{23}$ cm$^{-3}$\\
$\epsilon$ & Trapping parameter & 0.1 or 1.0 \\
$T_0$ & Initial temperature & $100-500$~K \\
$q$ & Radial dependence of temperature ($T\propto r^{-q}$) & 0 or 1 \\
$A_3$ & Flag to set reactions involving 3 active partners & on or off \\
  \hline    \hline
    \end{tabular}
\end{table*}

\begin{table*}
\caption{Results from the model described in the text. Results are shown for
the reactant species and are given as the log of the species abundance 
relative to H$_2$O.}
\label{tab:resultsa}
    \begin{tabular}{ccccccc}
\\
  \hline   \hline
Time (sec) & Density (cm$^{-3}$) & Temp (K) & CO & CH$_4$, HCN & 
H$_2$CO, CH$_3$OH & NH$_3$ \\  
     \hline
7.98$\times 10^{-11}$ & 7.89$\times 10^{20}$ & 462.1 & 
-0.86 & -1.43 & -1.55 & -2.03 \\
1.66$\times 10^{-10}$ & 6.26$\times 10^{20}$ & 427.7 &
-0.88 & -1.45 & -1.58 & -2.05 \\
2.59$\times 10^{-10}$ & 4.95$\times 10^{20}$ & 395.4 &
-0.89 & -1.46 & -1.59 & -2.07 \\
3.59$\times 10^{-10}$ & 3.94$\times 10^{20}$ & 366.6 & 
-0.90 & -1.47 &-1.60 &-2.08 \\
4.68$\times 10^{-10}$ & 3.09$\times 10^{20}$ & 337.9 &
-0.90 & -1.48 & -1.60 & -2.08 \\
5.85$\times 10^{-10}$ & 2.51$\times 10^{20}$ & 315.4 &
-0.91 & -1.48 & -1.61 & -2.09 \\
7.11$\times 10^{-10}$ & 1.99$\times 10^{20}$ & 292.1 &
-0.91 & -1.49 & -1.61 & -2.09 \\
8.48$\times 10^{-10}$ & 1.55$\times 10^{20}$ & 268.7 &
-0.91 & -1.49 & -1.61 & -2.09 \\
9.95$\times 10^{-10}$ &1.24$\times 10^{20}$ & 249.7 &
-0.91 & -1.49 & -1.61 & -2.09 \\
1.15$\times 10^{-9}$ & 9.40$\times 10^{19}$ & 227.3 &  
-0.91 & -1.49 & -1.61 & -2.09 \\
1.33$\times 10^{-9}$ & 7.78$\times 10^{19}$ & 213.4 &
-0.91 & -1.49 & -1.61 & -2.09 \\
1.51$\times 10^{-9}$ & 6.05$\times 10^{19}$ & 196.3 &
-0.91 & -1.49 & -1.61 & -2.09 \\
   \hline   \hline
    \end{tabular}
\end{table*}

\begin{table*}
\caption{Results from the model described in the text. Results are shown for
the product species and are given as the log of the species abundance 
relative to H$_2$O.}
\label{tab:resultsb}
    \begin{tabular}{ccccccccc}
\\
  \hline   \hline
Time (sec) & P$_1$, P$_2$ & P$_3$, P$_4$ & P$_5$ & P$_6$ &
P$_7$, P$_8$, P$_{10}$, P$_{11}$ & P$_9$, P$_{12}$ & P$_{13}$ &
P$_{14}$, P$_{15}$ \\
 \hline
7.98$\times 10^{-11}$ & -4.66 & -4.78 & -5.08 & -5.53 & -5.66 & -5.83 & 
-5.78 & -5.96 \\ 
1.66$\times 10^{-10}$ & -4.34 & -4.36 & -4.66 & -5.11 & -5.23 & -5.41 & 
-5.36 & -5.53 \\
2.59$\times 10^{-10}$ & -4.04 & -4.17 & -4.47 & -4.91 & -5.04 & -5.22 & 
-5.16 & -5.34 \\
3.59$\times 10^{-10}$ & -3.93 & -4.06 & -4.36 & -4.81 & -4.93 & -5.11 & 
-5.06 & -5.23 \\
4.68$\times 10^{-10}$ & -3.87 & -3.99 & -4.30 & -4.74 & -4.87 & -5.05 & 
-4.99 & -5.17 \\
5.85$\times 10^{-10}$ & -3.83 & -3.95 & -4.26 & -4.70 & -4.83 & -5.01 & 
-4.95 & -5.13 \\
7.11$\times 10^{-10}$ & -3.80 & -3.93 & -4.23 & -4.68 & -4.80 & -4.98 &
-4.92 & -5.10 \\
8.48$\times 10^{-10}$ & -3.78 & -3.91 & -4.21 & -4.66 & -4.78 & -4.96 & 
-4.91 & -5.09 \\
9.95$\times 10^{-10}$ & -3.77 & -3.90 & -4.20 & -4.65 & -4.77 & -4.95 &
-4.90 & -5.07 \\
1.15$\times 10^{-9}$ &  -3.76 & -3.89 & -4.19 & -4.64 & -4.76 & -4.94 &
-4.89 & -5.07 \\
      \hline
%
\\
Time (sec) & P$_{16}$ & P$_{17}$, P$_{18}$ & P$_{19}$, P$_{20}$ & P$_{21}$ &
P$_{22}$, P$_{23}$ & P$_{24}$, P$_{25}$ & P$_{26}$ & \\
 \hline
7.98$\times 10^{-11}$ & -2.05 & -2.62 & -2.74 & -3.22 & -3.47 & -3.59 & -4.07& 
\\
1.66$\times 10^{-10}$ & -1.84 & -2.41 & -2.53 & -3.02 & -3.27 & -3.40 & -3.87& 
\\
2.59$\times 10^{-10}$ & -1.74 & -2.31 & -2.44 & -2.92 & -3.19 & -3.31 & -3.79& 
\\
3.59$\times 10^{-10}$ & -1.69 & -2.26 & -2.38 & -2.87 & -3.14 & -3.27 & -3.75& 
\\
4.68$\times 10^{-10}$ & -1.66 & -2.23 & -2.35 & -2.84 & -3.12 & -3.24 & -3.72& 
\\
5.85$\times 10^{-10}$ & -1.64 & -2.21 & -2.33 & -2.82 & -3.10 & -3.22 & -3.70& 
\\
7.11$\times 10^{-10}$ & -1.62 & -2.20 & -2.32 & -2.81 & -3.09 & -3.21 & -3.69& 
\\
8.48$\times 10^{-10}$ & -1.62 & -2.19 & -2.31 & -2.80 & -3.08 & -3.21 & -3.68& 
\\
9.95$\times 10^{-10}$ & -1.61 & -2.18 & -2.31 & -2.79 & -3.08 & -3.20 & -3.68& 
\\
1.15$\times 10^{-9}$ &  -1.61 & -2.18 & -2.30 & -2.79 & -3.07 & -3.20 & -3.68& 
\\
 \hline     \hline
    \end{tabular}
\end{table*}

\begin{figure}
\epsscale{1.0}
\plotone{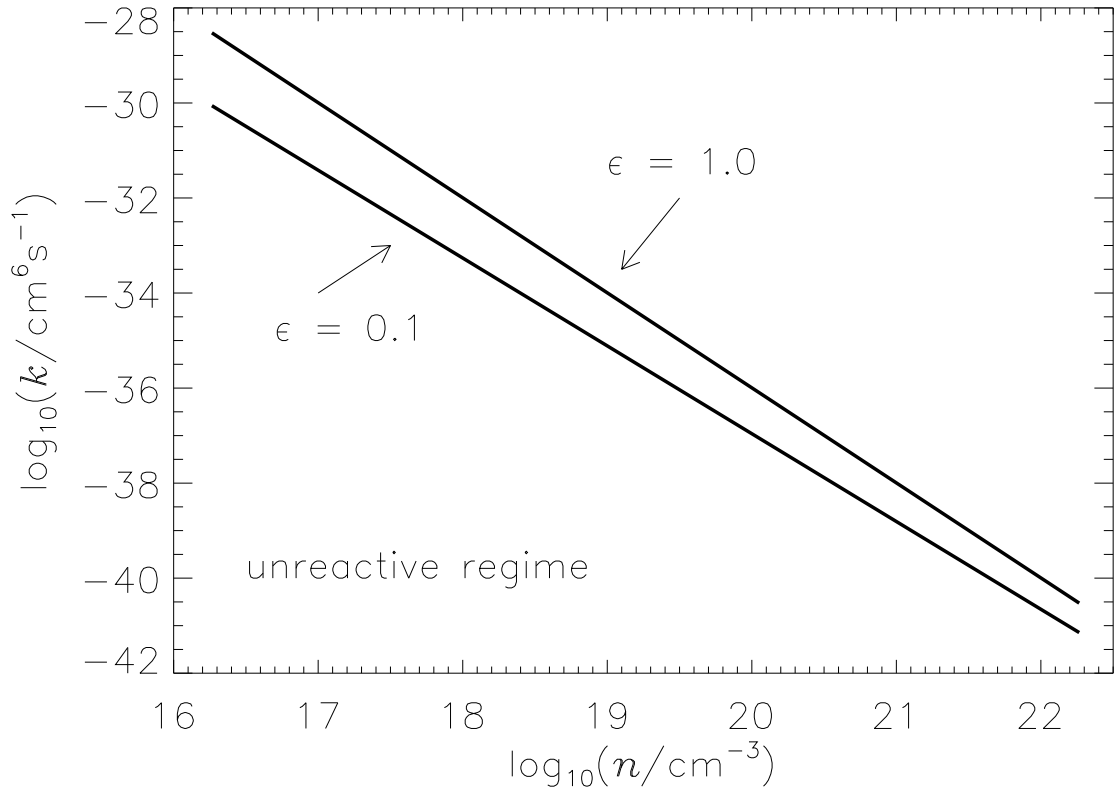}
\caption{The region above the lines indicate the values of $n$ and $k$ for 
which some products have abundances larger than $10^{-5}$ relative to H$_2$O, 
for a model in which $T_0 = 100$~K, $q = 0$, and $\alpha = 0$. When the 
expansion is partially hindered, $\epsilon = 0.1$, the chemistry is faster.}  
\label{fig1}
\end{figure}


\begin{thebibliography}{}
\bibitem[Allen \& Robinson(1977)]{AR77} Allen M. \& Robinson G.W., 1977, ApJ, 
212, 396
\bibitem[Belloche et al.(2008)]{B08} Belloche A., Menten K.M., Comito C., 
M\"{u}ller H.S.P., Schilke P., Ott J., Thorwirth S. \& Hieret C., 2008, A\&A,
482, 179
\bibitem[Belloche et al.(2009)]{B09} Belloche A., Garrod R.T., M\"{u}ller 
H.S.P., Menten K.M., Comito C. \& Schilke P., 2009, A\&A, 499, 215
\bibitem[Boogert et al.(2004)]{B04} Boogert et al., 2004, ApJS, 154, 359
\bibitem[Cecchi\textendash Pestellini et al.(2004)]{CCP04} Cecchi\textendash 
Pestellini C., Scappini F., Saija R., Iat\`{\i} M.A., Giusto A., Aiello S., 
Borghese F. \& Denti P., 2004, Int. J. Astrobiology, 3, 287
\bibitem[Collings et al.(2004)]{C04} Collings M.P., Anderson M.A., Chen R., 
Dever J.W., Viti S., Williams D.A. \& McCoustra M.R.S., 2004, MNRAS, 354, 1133
\bibitem[Duley(2000)]{D00} Duley W.W., 2000, MNRAS, 319, 791
\bibitem[Garrod, Widicus Weaver \& Herbst(2008)]{G08} Garrod R.T., Widicus
Weaver S.L. \& Herbst E., 2008, ApJ, 682, 283
\bibitem[Gibb et al.(2004)]{G04} Gibb E.L., Whittet D,C.B., Boogert A.C.A. \&
Tielens A.G.G.M., 2004, ApJS, 151, 35
\bibitem[Glasstone, Laidler \& Eyring(1941)]{G41} Glasstone S, Laidler K.J. \& 
Eyring H., 1946, Theory of Rate Processes, New York, McGraw\textendash Hill 
Book Company
\bibitem[Hasegawa, Herbst \& Leung(1992)]{H92} Hasegawa H., Herbst E. \&
Leung C.M., 1992, ApJS, 82, 167 
\bibitem[Herbst \& van Dishoeck(2009)]{HvD09} Herbst E. \& van Dishoeck E.F.,
2009, ARA\&A, 47, 427
\bibitem[Millar \& Hatchell(1998)]{MH98} Millar T.J. \& Hatchell J., 1998, 
Fa. Di., 109, 15
\bibitem[\"{O}berg et al.(2008)]{O08} \"{O}berg K.I., Garrod R.T., van Dishoeck
E.F. \& Linnartz H, 2009, A\&A, 504, 891
\bibitem[Snyder(2006)]{S06} Snyder E., 2006, Proc. Natl Ac. Sci., 103, 12243
\bibitem[Tielens \& Hagen(1982)]{TH82} Tielens A.G.G.M. \& Hagen W., 1982,
A\&A, 114, 245 
\bibitem[Tsuji(1973)]{T73} Tsuji T., 1973, A\&A, 23, 411
\bibitem[Viti et al.(2004)]{V04} Viti S., Collings M.P., Dever J.W., McCoustra 
M.R.S., Williams D.A., 2004, MNRAS, 354, 1141
\bibitem[Whittet(2002)]{W2nd} Whittet D.C.B., 2002, Dust in the Galactic 
Environment, 2nd edition, Institute of Physics Publishing, Bristol
\bibitem[Woodall et al.(2007)]{W07} Woodall J., Ag\'{u}ndez M., 
Markwick\textendash Kemper A.J. \& Millar T.J., 2007, A\&A, 466, 1197
\bibitem[Visscher et al.(2010)]{V10} Visscher C., Moses J.I. \& Saslow S.A., 
2010, Icarus, 209, 602
\end{thebibliography}
\end{document}